\documentstyle[prb,aps]{revtex}

\begin{document}

\draft

\preprint{\today}

\title{Global trends in the interplane penetration depth of layered superconductors}

\author{S.V.~Dordevic, E.J.~Singley and D.N.~Basov}
\address{Department of Physics, University of California,
San Diego, La Jolla, CA 92093}

\author{Seiki Komiya and Yoichi Ando}
\address{Central Research Institute of Electric Power Industry,
Tokyo, Japan}

\author{E.~Bucher$^*$}
\address{Lucent Technologies, Murray Hill, NJ 07974}

\author{C.C.~Homes and M.~Strongin}
\address{Department of Physics, Brookhaven National Laboratory,
Upton, NY 11973}

\wideabs{

\maketitle

\begin{abstract}
We report on generic trends in the behavior of the interlayer
penetration depth $\lambda_c$ of several  different classes of
quasi two-dimensional superconductors including high-T$_c$
cuprates, Sr$_2$RuO$_4$, transition metal dichalcogenides and
organic materials of the $(BEDT-TTF)_2X$ - series. Analysis of
these trends reveals two distinct patterns in the scaling between
the values of $\lambda_c$ and the magnitude of the c-axis DC
conductivity $\sigma_{DC}$: one realized in the systems with a
ground state formed out of well defined quasiparticles and the
other seen in systems in which the quasiparticles are not well
defined. The latter pattern is found primarily in under-doped
cuprates and indicates a dramatic enhancement (factor $\simeq
10^2$) of the energy scale $\Omega_C$ associated with the
formation of the condensate compared to the data for conventional
materials. We discuss the implication of these results on the
understanding of superconductivity in high-$T_c$ cuprates.
\end{abstract}
}

\narrowtext

\section{Introduction}

The formation of the superconducting  condensate in elemental
metals and their alloys  is well understood within the theory of
Bardeen, Cooper and Schrieffer (BCS) in terms  of a pairing
instability in the ensemble of Fermi liquid (FL) quasiparticles.
Applicability of the FL description to high-$T_c$ cuprate
superconductors is challenged by  remarkable  anomalies found in
both the spin- and charge response of these compounds in the
normal state \cite{orenstein00}. Because quasiparticles are not
well defined at $T>T_c$ in most cuprates it is natural to inquire
into the distinguishing characteristics of the superconducting
condensate which appears to be built from entirely different "raw
material". Infrared spectroscopy is perfectly suited for this
task. Indeed, the analysis of the optical constants in the far
infrared (IR) unfolds the process of the formation of the
condensate $\delta(0)$-peak in the dynamical conductivity
\cite{basov99} and also gives insight into the single-particle
excitations in the system both above and below $T_c$.

In this paper we focus on the inter-plane properties of high-$T_c$
superconductors. We will show that the distinctions in the
behavior of the condensate in conventional superconductors and
high-$T_c$ cuprates are most radical in the case of the $c$-axis
inter-plane response. The analysis of the generic trends seen in
the behavior of the $c$-axis condensate (correlation between the
penetration depth $\lambda_c$ and the DC conductivity
$\sigma_{DC}$) allows us to infer the energy scale $\Omega_C$
associated with the development of the superfluid in the cuprates.
This energy scale may dramatically exceed the energy gap in
systems lacking well defined quasiparticles at $T>T_c$ (primarily
in under-doped cuprates \cite{underdoped}). We discuss  a
connection between the magnitude of $\Omega_C$ and the nature of
the normal state response.

\section{Experimental Procedures}

The response of the superconducting condensate can be investigated
through the IR experiments probing the complex conductivity
$\sigma(\omega)=\sigma_1(\omega)+i\sigma_2(\omega)$ of a
superconductor. At $T<T_c$ the real part of the conductivity can
be written as:

\begin{equation}
\sigma_1^{SC}(\omega)=\frac{\rho_s}{8}\delta(0)+\sigma_1^{reg}(\omega).
\label{eq:definition}
\end{equation}
The delta-peak term represents the response of the condensate with
the superfluid density $\rho_s=4 \pi n_s e^2/m^*$ proportional to
the concentration of superconducting carriers $n_s$ and inversely
proportional to their effective mass $m^*$. The second term on the
right-hand side of the Eq.~\ref{eq:definition} is usually referred
to as the regular component and represents the conductivity that
is NOT due to the superconducting carriers. It may include
conductivity due to unpaired carriers at $T<T_c$ at finite
frequencies, phonons, interband transitions, magnons, etc.
Commonly, the condensate stiffness is characterized through the
penetration depth $\lambda=c/\sqrt{\rho_s}$, the notation we will
use in this paper.

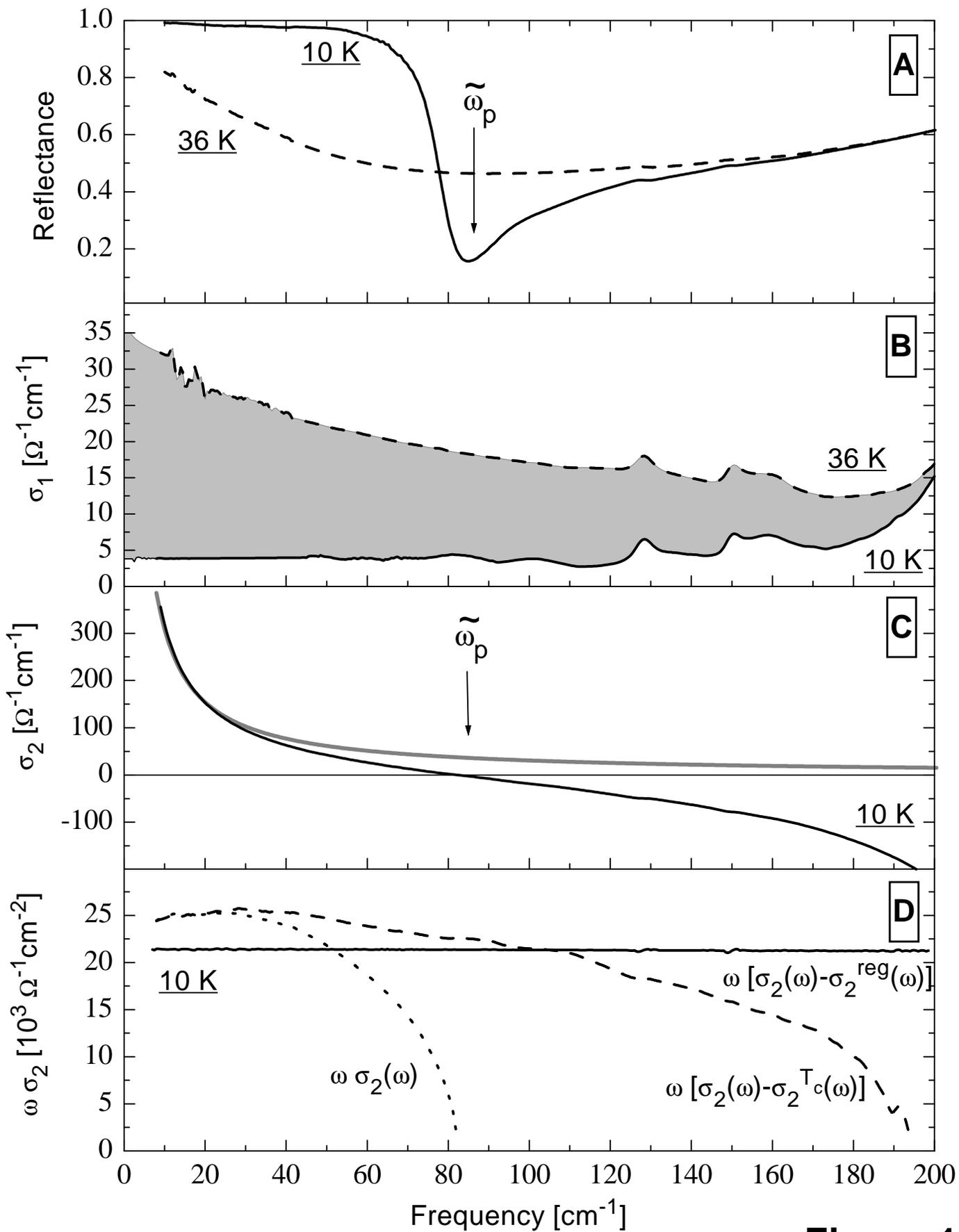
\begin{figure}
\caption{Interlayer response of LSCO single crystals with
T$_c$=36 K: reflectance R($\omega$) (panel A); real and imaginary
parts of the conductivity (panels B and C) and the product
$\sigma_2(\omega)\times\omega$ (panel D). The c-axis penetration
depth can be determined from the IR data using several different
techniques: from the position of the plasma minimum in
R($\omega$), from integrating the difference between the
$\sigma_1(\omega, T_c)$ and $\sigma_1(\omega, 10 K)$
(Eq.~\ref{eq:eq1}); and from examining the frequency dependence of
the $\sigma_2(\omega,10 K)\times \omega$. Advantages and
deficiencies of these methods are analyzed in Section II. The
latter approach may underestimate the magnitude of $\lambda_c$
because of the screening effects associated with the response of
unpaired charge carriers at T$\ll$T$_c$. We employed a
Kramers-Kronig transformation (Eqs.~\ref{eq:lambda} and
\ref{eq:sigma2}) to correct for this effect (solid line in panel
D).} \label{fig:four}
\end{figure}

In order to discuss several techniques that can be exploited to
determine the interlayer penetration depth of an anisotropic
superconductor we turn to our data for
La$_{1.83}$Sr$_{0.17}$CuO$_4$ (La214) with T$_c \simeq$ 36 K
(Fig.~\ref{fig:four}). Large single crystals were grown using
traveling-solvent floating-zone technique \cite{ando01} and were
carefully annealed to remove excess oxygen. The crystallographic
axes were determined by Laue diffraction and the samples were then
cut into platelets with the ac planes parallel to the wide face.
The error in the axes directions is less than $1^{\circ}$.
Near-normal-incidence reflectance measurements were performed at
UCSD in the frequency range between 10-48,000 cm$^{-1}$ (1 meV - 6
eV). The complex conductivity $\sigma(\omega)$ and complex
dielectric function
$\epsilon(\omega)=\epsilon_1(\omega)+i\epsilon_2(\omega_2)$ were
inferred from $R(\omega)$ using Kramers-Kronig (KK) analysis. The
low and high frequency extrapolations have negligible effect on
the data in the measured frequency interval.

Below we outline common analysis techniques used to determine the
penetration depth from the results of IR studies.

1) Raw c-axis reflectance of most  high-T$_c$ superconductors at
T$\ll$T$_c$ exhibits a sharp plasma edge. In the case of
La$_{1.83}$Sr$_{0.17}$CuO$_4$ this feature is located at $\sim$85
cm$^{-1}$ (Fig.~\ref{fig:four}, panel A). This behavior is in
contrast to the featureless normal state reflectance. The position
of the plasma edge is determined by the screened plasma frequency
$\tilde{\omega_p}$, from which the penetration depth can be
obtained as $c^2/\lambda^2 = \tilde{\omega_p}^2\epsilon_{\infty}$
(Ref.~\onlinecite{timusk89}). In the latter equation
$\epsilon_{\infty}$ is the real part of the dielectric constant
$\epsilon_1(\omega)$ at frequencies above the plasma edge. The
numerical value of $\epsilon_{\infty}$ is somewhat ambiguous and
introduces error in the result for $\lambda_c$. This technique has
been employed in
Ref.~\onlinecite{uchida97,motohashi00,singley01,tsvetkov98}.

2) In a BCS superconductor the formation of the condensate is
adequately described with the Ferrel-Glover-Tinkham (FGT) sum rule:
\begin{equation}
\rho_s=\frac{c^2}{\lambda^2}=\int_{0+}^{\Omega_C}[\sigma_1^{N}(\omega)-
\sigma_1^{reg}(\omega)]d \omega
\label{eq:eq1}
\end{equation}
where $\sigma_1^{N}(\omega)$ is the normal state conductivity at
T$_c$ and the upper integration limit $\Omega_C$ is of the order
of the gap energy. The upper cut-off issue for cuprates will be
discussed in detail below. According to this sum rule the area
"missing" from the normal state conductivity (shaded region in
Fig.~\ref{fig:four} B) is recovered under the $\delta$(0)-peak.
This technique may somewhat underestimate the magnitude of
$\lambda_c$ because, at least in under-doped cuprates the
superfluid density is accumulated from a broad energy region
significantly exceeding the gap energy
\cite{basov99,singley01,katz00,basov01}. This method has been used
for the analysis of the penetration depth in
Ref.~\onlinecite{bernhard99,schutzmann94,tajima97}.

3) Finally, the most commonly used method of extracting
$\lambda_c$ is based on the examination of the imaginary part of
the complex optical conductivity. By KK transformation, the
$\delta$-peak at $\omega=0$ in the real part of the optical
conductivity implies that the imaginary part has the form
$\sigma_2(\omega)=c^2/(4\pi\omega\lambda^2)$. Therefore the
magnitude of $\lambda_c$ can be estimated from
$\omega\times\sigma_2(\omega)$ in the limit of $\omega\rightarrow
0$. (the gray line in panel C of Fig.~\ref{fig:four} or the dotted
line in panel D)
(Ref.~\onlinecite{basov99,uchida97,singley01,katz00,basov01,basov94,uchida96,homes95,basov95}).

While method~3 is very well suited to quantify the magnitude of
the penetration depth, this technique also may introduce
systematic errors. Strictly speaking the relation
$\sigma_2(\omega)=c^2/(4\pi\omega\lambda^2)$ is valid only if
$\sigma_1^{reg}(\omega)=0$. Typically, this is not the case in
high-$T_c$ superconductors which all show residual absorption in
the far-IR conductivity. This absorption may be (in part)
connected with d-wave symmetry of the order parameter in cuprates
\cite{orenstein00} leading to gapless behavior at any finite
temperature. Data displayed in Fig.~\ref{fig:four} B clearly shows
non-vanishing IR conductivity down to the lowest T and $\omega$. A
finite regular contribution to $\sigma_1(\omega)$ implies a finite
contribution to $\sigma_2(\omega)$. Owing to this contribution the
spectra of $\sigma_2(\omega)$ acquire a complicated frequency
dependence that may significantly differ from the $1/\omega$ form
(Fig.~\ref{fig:four} C, D). Moreover, the magnitude of the
penetration depth extracted from such a spectrum is likely to be
underestimated, even if the product $\sigma_2(\omega)\times\omega$
is taken at the lowest experimentally accessible frequencies.

Systematic errors in the magnitude of $\lambda$ connected with
$\sigma_1^{reg}(\omega)>0$ can be eliminated using the following
procedure. The intrinsic value of the penetration depth can still
be determined from $\sigma_2(\omega)$, if the imaginary part of
the conductivity is corrected by $\sigma_2^{reg}(\omega)$
characterizing all screening effects that are not due to
superconducting carriers at $T<T_c$:

\begin{equation}
\sigma_2(\omega)-\sigma_2^{reg}(\omega)=
\frac{c^2}{4 \pi \omega \lambda^2}.
\label{eq:lambda}
\end{equation}
To determine $\sigma_2^{reg}(\omega)$ we employ a KK-like
transformation:
\begin{equation}
\sigma_2^{reg}(\omega)=-\frac{2 \omega}{\pi}\int_{0^+}^{\infty}
\frac{\sigma_1^{reg}(\omega')}{\omega'^2-\omega^2}d\omega'.
\label{eq:sigma2}
\end{equation}
The result of the application of the correction procedure for the
imaginary part of the conductivity is presented in
Fig.~\ref{fig:four} D. It appears that after subtraction of
$\sigma_2^{reg}(\omega)$ the remaining contribution to the
conductivity reveals a $1/\omega$ behavior over an extended
frequency region, supporting the soundness of the procedure
proposed here. We emphasize again that $no$ $other$ correction
procedure besides that described by Eqs.~\ref{eq:lambda} and
\ref{eq:sigma2} has been used. In the case of
La$_{1.83}$Sr$_{0.17}$CuO$_4$ the latter procedure has lead only
to minor correction of the absolute value of $\lambda_c$ ($\sim 18
\%$). That is because the absolute value of
$\sigma_1^{reg}(\omega)$ is relatively small and is constant
throughout far-IR (Fig.~\ref{fig:four} B). However, such a
correction can be much more significant for the overdoped samples
which often show stronger Drude-like contribution in
$\sigma_1^{reg}(\omega)$ spectra. Panel D also shows a frequently
used approximation to the method we have just outlined: instead of
subtracting $\sigma_2^{reg}(\omega)$, one subtracts
$\sigma_2(\omega, T_c)$ from $\sigma_2(\omega, T \ll T_c)$. The
resulting curve looks somewhat better than the uncorrected one,
but still yields an enhanced value of
$\omega\times\sigma_2(\omega)$ in the limit of $\omega\rightarrow
0$.

\section{Universal c-axis plot}

The $c$-axis penetration depth in a layered superconductor can be
determined from IR experiments
\cite{basov99,uchida97,motohashi00,singley01,tsvetkov98,katz00,basov01,bernhard99,schutzmann94,tajima97,basov94,uchida96,homes95,basov95}
as described in the previous section. In addition, several other
experimental techniques including magnetization measurements
\cite{trey73,huntley76,onabe78,carrington99,cooper90,taniguchi96,wanka96,yoshida96},
microwave absorption
\cite{pimenov00,shibauchi97,dressel94,prozorov00,pronin98,klein94,shibauchi94},
and vortex imaging \cite{kirtley98,moler98} can be used to
determine the magnitude of $\lambda_c$. Regardless of the method
employed, the interlayer penetration depth in several families of
cuprates reveals a universal scaling behavior with the magnitude
of $\sigma_{DC}(T=T_c)$ (Fig.~\ref{fig:basov})\cite{basov94}: the
absolute value of $\lambda_c$ is systematically suppressed with
the increase of the normal state conductivity \cite{errors}. The
scaling is obeyed primarily in under-doped cuprates (blue symbols
in Fig.~\ref{fig:basov}) . The deviations from the scaling are
also systematic and are most prominent in over-doped phases (red
symbols in Fig.~\ref{fig:basov}). Such deviations  are a direct
consequence of a well-established fact: on the over-doped side of
the phase diagram  $\sigma_{DC}$ increases whereas $\lambda_c$ is
either unchanged or may show a minor increase
\cite{katz00,uchida96,panagopoulos00}.

We find a similar scaling pattern between $\lambda_c$ and
$\sigma_{DC}$ in other classes of layered superconductors,
including organic materials, transition metal dichalcogenides and
Sr$_2$RuO$_4$ (Fig.~\ref{fig:basov}). While the non-cuprate data
set is not nearly as dense, the key trend is analogous to the one
found for cuprates. The slope of the $\lambda_c-\sigma_{DC}$
dependence is also close for both cuprates and non-cuprate
materials. The principal difference is that the cuprates universal
line is shifted down by approximately one order of magnitude in
$\lambda_c$. The latter result shows that the superfluid density
($\propto 1/\lambda^2$) is significantly enhanced in under-doped
cuprates compared to non-cuprate materials with the same DC
conductivity.

Possible origins of the $\lambda_c-\sigma_{DC}$ correlation were
recently discussed in the literature \cite{theory}. A plausible
qualitative account of this effect can be based on the FGT sum
rule, Eq.~\ref{eq:eq1}. For a dirty limit superconductor
$\sigma_1^N(\omega)\approx\sigma_{DC}$, and Eq.~\ref{eq:eq1} can
be approximated as:

\begin{equation}
\rho_s=\frac{c^2}{\lambda^2}\approx 2\Delta\sigma_{DC}.
\label{eq:eq2}
\end{equation}
Such an approximation is possible because within the BCS model the
energy scale $\Omega_C$ from which the condensate is collected is
of the order of magnitude of the gap: $\Omega_C\simeq
2\Delta\simeq 3-5 kT_c$. A connection between $1/\lambda^2$,
$\sigma_{DC}$, and $2\Delta$ is illustrated in the inset of
Fig.~\ref{fig:basov}. In the dirty limit the magnitude of
$\sigma_{DC}$ sets the amount of spectral weight available in the
normal state conductivity whereas the magnitude of $\Omega_C\simeq
2\Delta$ defines the fraction of this weight which is transferred
into condensate at $T<T_c$. Therefore, the magnitude of
$\lambda_c$ can be expected to systematically decrease with the
enhancement of the DC conductivity, in accord with the FGT sum
rule. Notably, an approximate form (Eq.~\ref{eq:eq2}) yields the
$\lambda_c-\sigma_{DC}$ scaling with the power law $\alpha=1/2$
which is close to $\alpha = 0.59$ seen in Fig.~\ref{fig:basov}.

The strong condensate density in the cuprates can be understood in
terms of the dramatic enhancement of the energy scale $\Omega_C$
over the magnitude of the energy gap. This can be seen through a
comparison of the universal scaling patterns observed for cuprates
and of a similar pattern detected for non-cuprate superconductors.
The energy scale associated with the condensate formation for
materials on the upper line, which for most conventional materials
in Fig.~\ref{fig:basov} is close to estimates of the gap, is of
the order of 1-3 meV. In Sr$_2$RuO$_4$ for example 2$\Delta$ = 2.2
meV based on Andreev reflection measurements. \cite{laube00} If
Eq.~\ref{eq:eq2}, in the form $\rho_s=c^2/\lambda^2\approx
\Omega_C \sigma_{DC}$ appropriate for cuprates, is employed to
describe the difference between the upper and the lower lines in
Fig.~\ref{fig:basov}, then one can conclude that the corresponding
scale for underdoped cuprates is $\sim$ 100 times greater, i.e. of
the order 0.1 - 0.3 eV. This assessment of $\Omega_C$ is supported
by the explicit sum rule analysis for several cuprates
\cite{basov99,basov01} and also makes $\Omega_C$ the largest
energy scale in the problem of cuprate superconductivity
\cite{summrulecomm}.

Data points in Fig.~\ref{fig:basov} for overdoped materials
support the notion that the $\lambda_c-\sigma_{DC}$ plot provides
means to learn about the energy scale associated with the
condensate formation. Deflection of the  over-doped cuprates from
the universal line implies that $\Omega_C$ is gradually suppressed
with the increased carrier density. This trend is common for
Tl$_2$Ba$_2$CuO$_{6+\delta}$ (Tl2201), La214 and
YBa$_2$Cu$_3$O$_{7-\delta}$ (YBCO) materials (see
Fig.~\ref{fig:basov}). Integration of the conductivity for all
these overdoped materials shows that the FGT sum rule is exhausted
at energies as low as 0.08 eV (Ref.~\onlinecite{katz00,basov01}).

In BCS superconductors $\Omega_C$ is related to $2\Delta$, and
therefore to T$_c$. In cuprates we find no obvious connection
between the broad energy scale $\Omega_C$ and the critical
temperature T$_c$. While scaling of $\lambda_c$ by the magnitude
of $T_c$ does reduce the "scattering" of the data points
\cite{shibauchi94,theory}, the two distinct $\lambda_c-
\sigma_{DC}$ patterns persist even if such scaling is implemented.
Similarly, the difference between the two lines in
Fig.~\ref{fig:basov} cannot be accounted for by differences in
T$_c$. In particular, the critical temperature of strongly
under-doped La214 materials is nearly the same as that of the
several ET-compounds ($\simeq 12-15$ K). Nevertheless, the
penetration depth is dramatically enhanced in the latter systems.

\section{In-plane quasiparticles and inter-plane transport}

A quick inspection of the materials in Fig.~\ref{fig:basov}
suggests that the $\simeq 3$ meV scale (top line) is observed in
systems in which superconductivity emerges out of a normal state
with well defined quasiparticles, whereas the enhanced value of
$\Omega_C\simeq$ 0.3 eV is found in underdoped cuprates for which
the quasiparticle concept may not apply (bottom line). The
experiments which in our opinion are most relevant to this
classification include quantum oscillations of the low-$T$
inter-layer resistivity (and of other quantities) in high magnetic
fields \cite{wosnitza96}. Quantum oscillations can be viewed as a
direct testimony of long-lived quasiparticles capable of
propagating coherently between the layers. Indeed, quantum
oscillations have been observed in 2D organic superconductors
\cite{wosnitza96,singleton01}, 2H-NbSe$_2$
(Ref.~\onlinecite{corcoran94}) and Sr$_2$RuO$_4$
(Ref.~\onlinecite{bergemann00}). On the contrary, quantum
oscillations have never been reported for under-doped cuprates.
The lack of coherence in the $c$-axis transport in these materials
indicates that the ground state of cuprates may be fundamentally
different.

Signatures of coherent and incoherent behavior can also be
recognized in the spectra of the $c$-axis conductivity. A hallmark
of a coherent response is the Drude peak seen in $\sigma_1(\omega)$
of metals. Notably, a similar feature has never been found in
underdoped compounds (forming the lower line in
Fig.~\ref{fig:basov}). The electronic contribution to
$\sigma_1(\omega)$ in these materials is usually structureless
which is commonly associated with the incoherent (diffusive)
motion of charge carriers across the planes. On the contrary, many
materials that belong to the upper line in Fig.~\ref{fig:basov}
demonstrate a familiar Drude-like behavior. This kind of behavior
has been found in Sr$_2$RuO$_4$ (Ref.\onlinecite{katsufuji96}) and
is also shown in our data \cite{dordevic01} for inter-plane
response of 2H-NbSe$_2$ (Fig.~\ref{fig:six}, top-right panel). In
both cases, the width of the peak decreases at low temperatures,
which is characteristic of the response of ordinary metals
\cite{kennedy84}. As for the over-doped cuprates (located in a
cross-over region between the two lines in Fig.~\ref{fig:basov})
their conductivity is indicative of the formation of the
Drude-like peak (see for example $\sigma_c(\omega)$ for
YBa$_2$Cu$_3$O$_{7}$; Fig.~\ref{fig:six}, top-middle panel), which
is becoming more pronounced with increased carrier density
\cite{commdrude}.

Analysis of the anisotropic carrier dynamics in several layered
superconductors indicates that the degree of coherence in the
interplane transport may be related to the strength of inelastic
scattering within the conducting planes. The bottom panels in
Fig.~\ref{fig:six} show the $in-plane$ scattering rate (inverse
lifetime) $1/\tau_{ab}(\omega)$ (Ref.~\onlinecite{tau}) for the
layered compounds corresponding to the top three panels
\cite{ybcocomm}. In all these systems
$1/\tau_{ab}(\omega)\propto\omega$ over an extended frequency
interval (up to 3,000 cm$^{-1}$) \cite{lineartau}. An important
feature of the data displayed in Fig.~\ref{fig:six} is that as
doping is increased from underdoped YBa$_2$Cu$_3$O$_{6.6}$ to
optimally doped YBa$_2$Cu$_3$O$_{6.95}$ the absolute values of
$1/\tau_{ab}(\omega)$ decrease. A similar trend is observed in
other cuprate families
\cite{basov96,puchkov96b,rotter91,puchkov96}. The shaded regions
in Fig.~\ref{fig:six} represent Landau Fermi liquid (LFL) regime,
where the quasiparticles are well defined, i.e. the magnitude of
the scattering rate is smaller than energy
($1/\tau(\omega)\leq\omega$). In 2H-NbSe$_2$ $1/\tau_{ab}(\omega,
10 K)$ is in the LFL regime over the entire frequency interval
displayed in Fig.~\ref{fig:six}. However this is not the case for
the two cuprates discussed. We believe that these differences in
absolute values may have a profound effect on the $interplane$
transport. In 2H-NbSe$_2$ where the in-plane quasiparticles are
well defined the interplane transport is also coherent, and is
characterized by a narrow Drude-like mode whose width decreases
with temperature (Fig.~\ref{fig:six}, top-right panel). On the
other hand, in YBa$_2$Cu$_3$O$_{6.6}$, which is lacking well
defined quasiparticles, the interplane transport is incoherent,
with $\sigma_1(\omega)$ being dominated by optical phonons
(Fig.~\ref{fig:six}, top-right). As for the over-doped
YBa$_2$Cu$_3$O$_{7}$ (Fig.~\ref{fig:six}, bottom-middle) the
optical conductivity of this compound is in between these two
opposite limits. Figure~\ref{fig:six} therefore supports the
notion that long-lived $in-plane$ quasiparticles may be one of the
necessary prerequisites for coherent $out-of-plane$ transport.

\section{Global Trends in Layered Superconductors}

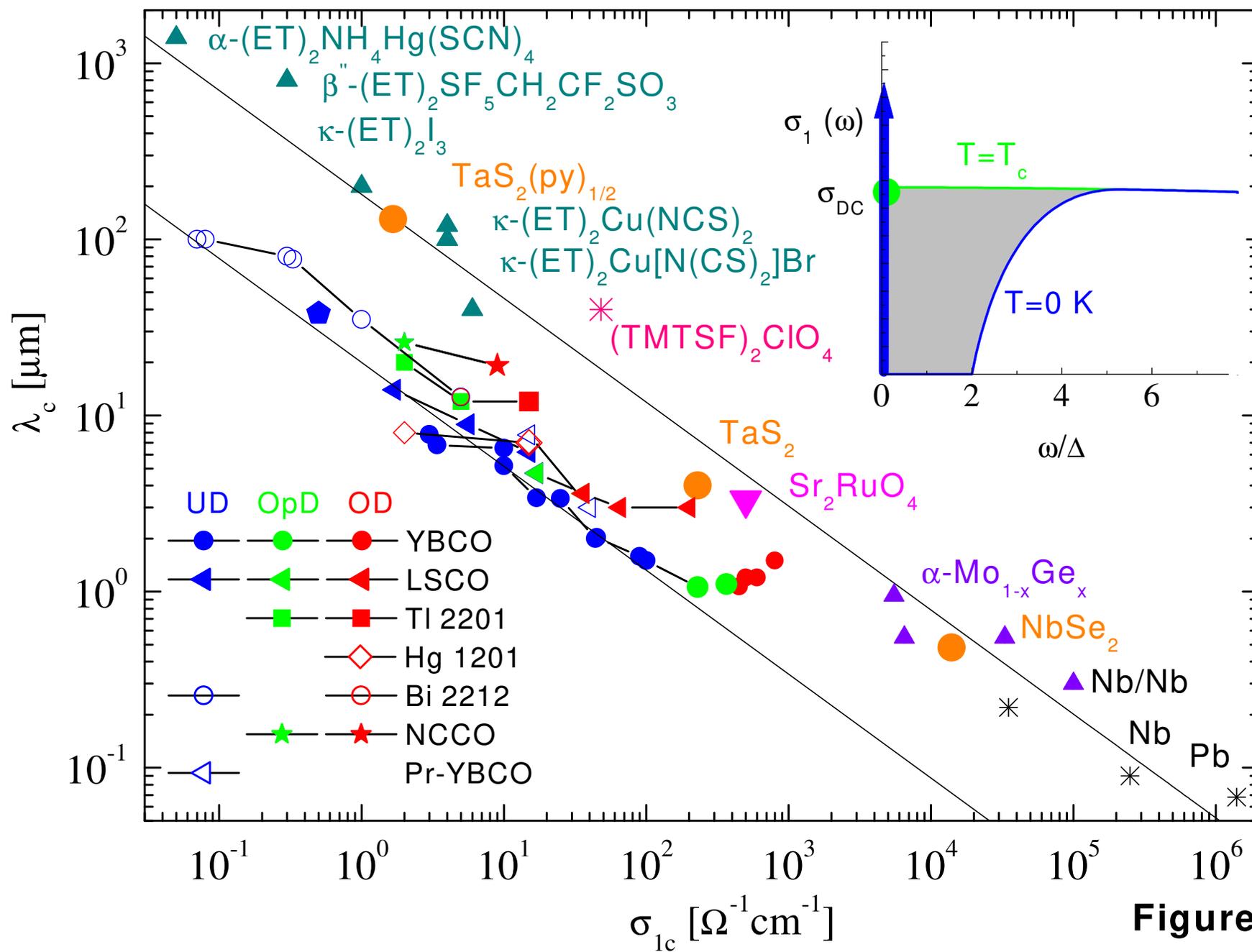
\begin{figure}
\caption{The $c$-axis penetration depth $\lambda_c(T=0 K)$ as a
function of the $c$-axis DC conductivity $\sigma_{DC}(T_c)$. We
find two distinct patterns of $\lambda_c - \sigma_{DC}$ scaling.
Cuprate superconductors exhibit much shorter penetration depths
than non-cuprates materials with the same $\sigma_{DC}(T_c)$. This
result implies a dramatic enhancement of the energy scale
$\Omega_C$ from which the condensate is collected as described in
the text. The superconducting transition temperature T$_c$ has not
been found to be relevant to the $\lambda_c - \sigma_{DC}$
scaling. Data points: YBCO
(Ref.~\protect\onlinecite{uchida97,basov94,homes95}), overdoped
YBCO
(Ref.~\protect\onlinecite{bernhard99,schutzmann94,tajima97,basov94}),
La214
(Ref.~\protect\onlinecite{uchida97,uchida96,basov95,shibauchi94}),
HgBa$_2$Cu$_2$O$_4$ (Ref.~\protect\onlinecite{kirtley98,singley}),
Tl2201 (Ref.~\protect\onlinecite{basov99,katz00,moler98}),
Bi$_2$Sr$_2$CaCu$_2$O$_8$
(Ref.~\protect\onlinecite{motohashi00,cooper90}),
Nd$_{2-x}$Ce$_x$CuO$_4$
(Ref.~\protect\onlinecite{singley01,pimenov00}). Blue points -
underdoped (UD), green - optimally doped (OpD); red - overdoped
(OD). Transition metal dichalcogenides
(Ref.~\protect\onlinecite{trey73,huntley76,onabe78,kennedy84,garoche76,finley80,thompson72}),
(ET)$_2$X compounds
(Ref.~\protect\onlinecite{carrington99,taniguchi96,wanka96,shibauchi97,dressel94,prozorov00,su98,kajita87,su99}),
(TMTSF)$_2$ClO$_4$ (Ref.~\protect\onlinecite{schwenk83,murata81}),
Sr$_2$RuO$_4$ (Ref.~\protect\onlinecite{yoshida96,maeno94}),
niobium (Ref.~\protect\onlinecite{pronin98,klein94}), lead
(Ref.~\protect\onlinecite{klein94}), niobium Josephson junctions
(Ref.~\protect\onlinecite{goldobin98}) and
$\alpha$Mo$_{1-x}$Ge$_{x}$
(Ref.~\protect\onlinecite{lemberger00}). Inset: in a conventional
dirty limit superconductor the spectral weight of the
superconducting condensate (given by $1/\lambda^2$) is collected
primarily from the energy gap region. The total normal weight is
preset by magnitude of $\sigma_{DC}$ whereas the product of
$2\Delta\times \sigma_{DC}$ quantifies the fraction of the weight
that condenses.} \label{fig:basov}
\end{figure}

To summarize the experimental results reported in this work, we
wish to stress the following points: {\it i)} two distinct
patterns in $\lambda_c-\sigma_{DC}$ correlation
(Fig.~\ref{fig:basov}) are indicative of a dramatic difference
($\simeq 10^2$) in the energy scale $\Omega_C$ from which the
interlayer condensate is collected; {\it ii)} the pattern with the
typical energy scale of $\Omega_C \simeq 3$ meV is realized in the
materials with the coherent transport between the planes, whereas
the one with $\Omega_C \simeq 300$ meV is found in underdoped
cuprate superconductors with an incoherent response; {\it iii)}
over-doped cuprates reveal a cross-over between the two behaviors;
{\it iv}) coherence in the interlayer transport correlates with
the strength of inelastic scattering within the conducting planes
(Fig.~\ref{fig:six}).

These results allow us to draw several conclusions regarding
features of the superconducting condensate in different layered
systems:
\\
 $\bullet$ The symmetry of the order parameter seems to be unrelated
to trends seen in the c-axis condensate response. Indeed, the upper line
in Fig.~\ref{fig:basov} is formed by  $s$-wave transition metal
dichalcogenides, $p$-wave Sr$_2$RuO$_4$ and organic materials for
which both $s$- and $d$-wave states have been proposed
\cite{wosnitza96}, while $d$-wave high-$T_c$ materials form the
lower line and the crossover region between the lines. \\
 $\bullet$  Electrodynamics of the systems on the top line at $T\ll T_c$
is determined by the magnitude of the gap (and hence by $T_c$), in general
agreement with the BCS theory. It is therefore hardly surprising
that the trend initiated by 2D superconductors is also followed
in 1-dimensional organic conductors, as well as by more
conventional systems such as Nb Josephson junctions, bulk Nb and Pb or
amorphous $\alpha$Mo$_{1-x}$Ge$_x$ (see Fig.~\ref{fig:basov}).\\
 $\bullet$ While the pseudogap state has been shown to be responsible for
the anomalous superfluid response of the underdoped cuprates
\cite{basov99,basov01}, the characteristic pseudogap temperature
$T^*=90- 350$ K is still much lower than our estimate of $\Omega_C$
for these materials (0.1 - 0.3 eV, i.e. 1,000 - 3,000 K).\\
$\bullet$ Unlike BCS superconductors where T$_c$ is determined by
$2\Delta$ and therefore by $\Omega_C$, the critical temperature
T$_c$ in cuprates correlates with neither $2\Delta$ nor $\Omega_C$.

In conclusion, analyzing a large amount of experimental data, we
found two distinctly different patterns in $\lambda_c-\sigma_{DC}$
scaling in layered superconductors. Based on the universal c-axis
plot, we inferred a broad energy scale $\Omega_c$ relevant for
pair formation in underdoped cuprates. This result is consistent
with the idea that the superconducting transition in the cuprates
is driven by a lowering of the electronic kinetic energy
\cite{norman}. We argue that the appearance of such an energy
scale is fundamentally related to the incoherent c-axis transport,
which on the other hand may be related to poorly defined in-plane
quasiparticles. A quantitative account of the distinct energy
scales associated with the condensate is a challenge for models
attempting to solve the puzzle of cuprate superconductivity.

The research at UCSD is supported by NSF, DoE and the Research
Corporation. The work at Brookhaven National Laboratory is
supported, in part, by the U.S. Department of Energy, Division of
Materials Science, under Contract No. DE-AC02-98CH10886.

* Also at Fakult\"{a}t f\"{u}r Physik, Universit\"{a}t Konstanz,
D-78457 Konstanz, Germany.

Electronic address: sasa@physics.ucsd.edu

\begin{figure}
\caption{Examples of the interplane transport for layered
superconductors. Top panels show the out-of-plane optical
conductivity $\sigma_c(\omega)$, the bottom panels the
corresponding in-plane scattering rate $1/\tau_{ab}(\omega)$. The
observation of the Drude-like feature in the interplane optical
conductivity of the dichalcogenide 2H-NbSe$_2$ (top-right panel)
is consistent with magnetoresistance measurements that revealed
evidence for well-behaved quasiparticles. In contrast the
conductivity of underdoped YBa$_2$Cu$_3$O$_{6.6}$ material
(top-left panel) gives no signs of coherent response. Overdoped
cuprates show the emergence of a Drude-like feature (top-middle
panel) and also occupy an intermediate position between the two
lines in Fig.~\ref{fig:basov}. Experimental data:
YBa$_2$Cu$_3$O$_{6.6}$
(Ref.~\protect\onlinecite{basov96,homes93}), YBa$_2$Cu$_3$O$_{7}$
(Ref.~\protect\onlinecite{schutzmann94,tajima97}),
YBa$_2$Cu$_3$O$_{6.95}$ (Ref.~\protect\onlinecite{basov96}) and
2H-NbSe$_2$ (Ref.~\protect\onlinecite{dordevic01}).}
\label{fig:six}
\end{figure}


\begin{references}

\bibitem{orenstein00} J.~Orenstein and A.J.~Millis, Science {\bf 288},
468 (2000).

\bibitem{basov99} D.N.~Basov, S.I.~Woods, A.S.~Katz, E.J.~Singley,
R.C.~Dynes, M.~Xu, D.G.~Hinks, C.C.~Homes and M.~Strongin, Science
{\bf 283}, 49 (1999). 

\bibitem{underdoped} Under-doped cuprates are compounds whose
carrier doping level is smaller than optimal, i.e. the one that gives the
highest T$_c$.

\bibitem{ando01} Y.~Ando, A.N.~Lavrov, S.~Komiya, K.~Segawa and
X.F.~Sun, Phys.Rev.Lett. {\bf 87}, 017001 (2001).

\bibitem{timusk89} T.~Timusk and D.B.~Tunner, in {\it Physical
properties of high temperature superconductors I}, edited by
D.M.~Ginsberg, World Scientific (1989).

\bibitem{uchida97} S.~Uchida and K.~Tamasaku, Phys.C {\bf 293},
1 (1997). 

\bibitem{motohashi00} T.~Motohashi, J.~Shimoyama, K.~Kitazawa, K.~Kishio,
K.M.~Kojima, S.~Uchida and S.~Tajima, Phys.Rev.B {\bf 61},
9269 (2000). 

\bibitem{singley01} E.J.~Singley, D.N.~Basov, K.~Kurahashi, T.~Uefuji and
K.~Yamada, Phys.Rev.B {\bf 64}, 224503 (2001). 

\bibitem{tsvetkov98} A.A.~Tsvetkov, D.~van der Marel,
K.A.~Moler, J.R.~Kirtley, J.L.~de Boer, A.~Meetsma, Z.F.~Ren,
N.~Koleshnikov, D.~Dulic, A.~Damascelli, M.~Gruninger,
J.~Schutzmann, J.W.~van der Eb, H.S.~Somal, J.H.~Wang, Nature,
{\bf 395}, 360 (1998).

\bibitem{katz00} A.S.~Katz, S.I.~Woods, E.J.~Singley,
T.W.~Li, M.~Xu, D.G.~Hinks, R.C.~Dynes and D.N.~Basov,
Phys.Rev.B {\bf 61}, 5930 (2000). 

\bibitem{basov01} D.N.Basov, C.C.~Homes, E.J.~Singley, M.~Strongin,
T.~Timusk, G.~Blumberg and D.~van der Marel, Phys.Rev.B {\bf 63},
134514 (2001);

\bibitem{bernhard99} C.~Bernhard, D.~Munzar, A.~Wittlin, W.~Konig,
A.~Golnik, C.T.~Lin, M.~Klaser, T.~Wolf, G.~Muller-Vogt and
M.~Cardona, Phys.Rev.B {\bf 59}, 6631 (1999)

\bibitem{schutzmann94} J.~Schutzmann, S.~Tajima, S.~Miyamoto and
S.~Tanaka, Phys.Rev.Lett. {\bf 73}, 174 (1994). 

\bibitem{tajima97} S.~Tajima, J.~Schutzmann, S.~Miyamoto, I.~Terasaki,
Y.~Sato and R.~Hauff, Phys.Rev.B {\bf 55},
6051 (1997). 

\bibitem{basov94} D.N.~Basov, T.~Timusk, B.~Dabrowski and J.D.~Jorgensen,
Phys.Rev.B {\bf 50}, 3511 (1994). 

\bibitem{uchida96} S.~Uchida, K.~Tamasaku and S.~Tajima, Phys.Rev.B {\bf
53}, 14558 (1996). 

\bibitem{homes95} C.C.Homes, T.~Timusk, D.A.~Bonn, R.~Liang and
W.N.~Hardy, Phys.C {\bf 254}, 265 (1995). 

\bibitem{basov95} D.N.~Basov, H.A.~Mook, B.~Dabrowski  and T.~Timusk,
Phys.Rev.B {\bf 52}, 13141 (1995). 

\bibitem{trey73} P.~de Trey, S.~Gygax and J.-P.~Jan, J.Low Temp.Phys,
{\bf 11}, 421 (1973). 

\bibitem{huntley76} D.J.~Huntley and R.F.~Frindt, in {\it Physics and
chemistry of materials with layered structures}, Reidel Publishing
Company, 1976. 

\bibitem{onabe78} K.~Onabe, M.~Naito and S.~Tanaka, J.Phys.Soc.Jap.
{\bf 45} 50 (1978). 

\bibitem{carrington99} A.~Carrington, I.J.~Bonalde, R.~Prozorov,
R.W.~Giannetta, A.M.~Kini, J.~Schlueter, H.H.~Wang, U.~Geiser and
J.M.~Williams, Phys.Rev.Let. {\bf 83},
4172 (1999). 

\bibitem{cooper90} J.R.~Cooper, L.~Forro and B.~Keszei,
Nature {\bf 343}, 444 (1990). 

\bibitem{taniguchi96} H.~Taniguchi, H.~Sato, Y.~Nakazawa and K.~Kanoda,
Phys.Rev. B {\bf 53}, 8879 (1996).

\bibitem{wanka96} S.~Wanka, D.~Beckmann, J.~Wosnitza, E.~Balthes,
D.~Schweitzer, W.~Strunz and H.J.~Keller, Phys.Rev. B {\bf 53},
9301 (1996). 

\bibitem{yoshida96} K.~Yoshida, Y.~Maeno, S.~Nishizaki and
T.~Fujita, Phys.C {\bf 263},
519 (1996). 

\bibitem{pimenov00} A.~Pimenov, A.V.~Pronin, A.~Loidl,
U.~Michelucci, A.P.~Kampf, S.I.~Krasnosvobodtsev, V.S.~Nozdrin,
D.~Rainer, Phys.Rev.B {\bf 62}, 9822 (2000). 

\bibitem{shibauchi97} T.~Shibauchi, M.~Sato, A.~Mashio, T.~Tamegai,
H.~Mori, S.~Tajima and S.~Tanaka, Phys.Rev. B {\bf 55}, 11977
(1997). 

\bibitem{dressel94} M.~Dressel, O.~Klein, G.~Gruner, K.D.~Carlson,
H.H.~Wang and  J.M.~Williams, Phys.Rev. B {\bf 50},
13603 (1994). 

\bibitem{prozorov00} R.~Prozorov, R.W.~Giannetta, J.~Schlueter,
A.M.~Kini, J.~Mohtasham, R.W.~Winter and G.L.~Gard,
Phys.Rev.B {\bf 63}, 052506 (2001). 

\bibitem{pronin98} A.V.~Pronin, M.~Dressel, A.~Pimenov, A.~Loidl,
I.V.~Roshchin and L.H.~Greene, Phys.Rev.B {\bf 57}, 14416
(1998). 

\bibitem{klein94} O.~Klein, E.J.~Nicol, K.~Holczer and G.~Gruner,
Phys.Rev.B {\bf 50}, 6307 (1994). 

\bibitem{shibauchi94} T.~Shibauchi, H.~Kitano, K.~Uchinokura, A.~Maeda,
T.~Kimura and K.~Kishio, Phys.Rev.Lett. {\bf 72},
2263 (1994). 

\bibitem{kirtley98} J.R.~Kirtley, K.A.~Moler, G.~Villard and
A. Maignan, Phys.Rev.Lett. {\bf 81}, 2140 (1998). 

\bibitem{moler98} K.A.~Moler, J.R.~Kirtley, D.G.~Hinks, T.W.~Li and M.~Xu,
Science {\bf 279}, 1193 (1998);  

\bibitem{errors} Whenever available $\lambda_c$ and $\sigma_{DC}$
values obtained from IR spectroscopy were used, and as shown here
the errors of such measurements do not exceed 15-20 $\%$. This
error includes the error introduced by the uncertanity of
reflectance measurements, which is typically around 1 $\%$. When
IR spectroscopic data were not available (particularly for the
compounds on the top line in Fig.~\ref{fig:basov}), we used the
results from other experimental techniques mentioned in the text.
Typical errors, as reported in the original publications, are:
$<$10 $\%$ for microwave absorption \cite{pronin98}, $\sim$20 $\%$
for magnetization \cite{carrington99}, $<$30 $\%$ for vortex
imaging \cite{moler98} and $\sim$ 5 $\%$ for resistivity (i.e.
1/$\sigma_{DC}$) measurements. When the results from different
measurements on the same compound differed substantially, for
example $\lambda_c$ values for $\kappa (BEDT-TTF)_2Cu[N(CS)_2]Br$
(Ref.~\onlinecite{carrington99,dressel94}), they were both shown
in Fig.~\ref{fig:basov}. We emphasize that errors as high as 20
$\%$ cannot in any significant way change a plot that spans over
several orders of magnitude.

\bibitem{panagopoulos00} C.~Panagopoulos, J.R.~Cooper, T.~Xiang, Y.S.~Wang
and C.W.~Chu, Phys.Rev.B, {\bf 61}, 3808 (2000). 

\bibitem{theory} W.Kim and J.P.~Carbotte, Phys.Rev. B {\bf 61},
11886 (2000); Y.Ohashi, J.Phys.Soc.Jpn {\bf 69}, 659 (2000);
P.J.~Hirschfeld, S.M.~Quinlan and D.J.~Scalapino, Phys.Rev. B {\bf
55}, 12742 (1997); S.~Chakravarty, Hae-Young Kee and E.~Abrahams,
Phys.Rev.Lett. {\bf 82}, 2366, (1999).

\bibitem{laube00} F.~Laube, G.~Goll, H.v.~Lohneysen, M.~Fogelstrom,
and F.~Lichtenberg, Phys.Rev.Lett. {\bf 84}, 1595 (2000).

\bibitem{summrulecomm}
The sum rule arguments discussed in this section may help one
understand the rationale behind the universal $\lambda_c -
\sigma_{DC}$ scaling in underdoped cuprates. That is despite the
fact that the superconducting energy gap is not well-defined in
the interlayer conductivity of these materials. The gap-less
response of cuprates is exemplified in the top-left panel of
Fig.~\ref{fig:six} displaying $\sigma_1(\omega)$ data for
YBa$_2$Cu$_3$O$_{6.6}$ with $T_c\simeq 60$ K. The normal state
conductivity is suppressed as the sample is cooled down to $T_c$,
with a transfer of spectral weight to higher energies. Below
$T_c$, one does not find any radical changes in the
$\sigma_1(\omega)$ spectra. Most importantly, this system, along
with all other under-doped compounds, shows significant absorption
in the superconducting state so that $\sigma_1^{reg} > 0$.
Therefore, in cuprates only a {\it small fraction} of the
far-infrared spectral weight is contributing to the condensate.
The latter result is in apparent conflict with the assumption
$\sigma_1^{reg}(\omega<2\Delta)\simeq 0$, which allows one to
reduce Eq.~\ref{eq:eq1} to an approximate form given by
Eq.~\ref{eq:eq2}. However, the strong condensate density seen in
the cuprates located on the universal line can be understood in
terms of the dramatic enhancement of the energy scale $\Omega_C$
over  the magnitude  of the energy gap. This latter conclusion
also follows from the explicit sum rule analysis for samples of
underdoped La$_{2-x}$Sr$_x$CuO$_4$ (La214),
YBa$_2$Cu$_3$O$_{7-\delta}$ (YBCO) and
Tl$_2$Ba$_2$CuO$_{6+\delta}$ (Tl2201) materials, suggesting that
$\Omega_C$ in these compounds exceeds 0.1-0.2 eV
(Ref.~\onlinecite{basov99,katz00,basov01}).

\bibitem{wosnitza96} J.~Wosnitza, {\it Fermi surfaces of low-dimensional
organic metals and superconductors}, Berlin, New York; Springer
(1996).

\bibitem{singleton01} J.~Singleton, P.A.~Goddard, A.~Ardavan,
N.~Harrison, S.J.~Blundell, J.A.~Schlueter and A.M.~Kini,
Phys.Rev.Lett. {\bf 87}, 117001 (2001). 

\bibitem{corcoran94} R.~Corcoran, P.~Meeson, Y.~Onuki, P.-A.~Probst,
M.~Springford, K.~Takita, H.~Harima, G.Y.~Guo, B.L.~Gyorffy,
J.Phys.Cond.Mat. {\bf 6}, 4479 (1994). 

\bibitem{bergemann00} C.~Bergemann, S.R.~Julian, A.P.~Mackenzie,
S.~NishiZaki and Y. Maeno, Phys.Rev.Lett. {\bf 84},
2662 (2000). 

\bibitem{katsufuji96} T.~Katsufuji, M.~Kasai, Y.~Tokura,
Phys.Rev.Lett. {\bf 76}, 126 (1996). 

\bibitem{dordevic01} S.V.~Dordevic, D.N.~Basov, R.C.~Dynes and E.~Bucher,
Phys.Rev.B {\bf 64}, 161103 (2001). 

\bibitem{kennedy84} Earlier measurements of the $c$-axis
infrared absorption proved applicability of conventional BCS
electrodynamics to the 2H-NbSe$_2$ data. See: R.J.~Kennedy and
B.P.~Clayman, Can.J.Phys. {\bf 62}, 776 (1984). 

\bibitem{commdrude}
We emphasize that the observation of a Drude-like mode in the
optical spectra is consistent with, but is not a definite proof of
coherent interlayer transport. Similarly, the absence of a
zero-energy mode, does not necessarily imply incoherent transport.
Recent c-axis IR experiments on 2D organic superconductors
\cite{mcguire01} seem to indicate the absence of a Drude-like mode
in these compounds. However, high magnetic field measurements
\cite{wosnitza96,singleton01} undoubtedly show the existence of 3D
Fermi surface, i.e. well defined quasiparticles that can propagate
coherently between the layers.

\bibitem{tau} The scattering rate can be calculated from
IR data as:
\begin{equation}
\frac{1}{\tau_{ab}(\omega)}=\frac{\omega_{p}^{2}}{4 \pi}
\frac{\sigma_1(\omega)} {\sigma_1^2(\omega)+\sigma_2^2(\omega)},
\end{equation}
where $\omega_p$ is the plasma frequency and can be obtained from
the integration of the optical conductivity $\sigma_{1}(\omega)$
up to the frequency corresponding to the onset of interband
absorption. See Ref.~\onlinecite{puchkov96} for a review.

\bibitem{ybcocomm} Because the in-plane measurements of
YBa$_2$Cu$_3$O$_{7}$ are not available, we have used
$1/\tau_{ab}(\omega)$ data for YBa$_2$Cu$_3$O$_{6.95}$. We believe
that this does not affect the conclusions of the paper in any
significant way.

\bibitem{lineartau} Many other 2D conductors also show a linear
frequency dependence of the $in-plane$ scattering rate. In particular
this behavior has been reported for graphite \cite{xu96}, 2H-TaSe$_2$
and other transition metal dichalcogenides \cite{valla00}, and a
variety of cuprates \cite{basov96,puchkov96b,rotter91,puchkov96}.

\bibitem{basov96} D.N.~Basov, R.~Liang, B.~Dabrowski, D.A.~Bonn,
W.N.~Hardy and T. Timusk, Phys.Rev.Lett. {\bf 77}, 4090 (1996).

\bibitem{puchkov96b} A.V.~Puchkov, P.~Fournier, D.N.~Basov, T.~Timusk,
A.~Kapitulnik and N.N.~Kolesnikov, Phys.Rev.Lett. {\bf 77}, 3212
(1996).

\bibitem{rotter91} L.D.~Rotter, Z.~Schlesinger, R.T.~Collins,
F.~Holtzberg, C.~Field, U.W.~Welp, G.W.~Crabtree, J.Z.~Liu,
Y.~Fang, K.G.~Vandervoort and S.~Fleshler, Phys.Rev.Lett. {\bf
67}, 2741 (1991).

\bibitem{puchkov96} A.V.~Puchkov, D.N.~Basov and T.~Timusk,
J.Phys.Cond.Mat. {\bf 8}, 10049 (1996).

\bibitem{mcguire01} J.J.~McGuire, T.~Room, A.~Pronin, T.~Timusk,
J.A.~Schlueter, M.E.~Kelly and A.M.~Kini,
Phys.Rev.B {\bf 64}, 94503 (2001). 

\bibitem{xu96} S.~Xu, J.~Cao, C.C.~Miller, D.A.~Mantell, R.J.D.~Miller
and Y.~Gao, Phys.Rev.Lett. {\bf 76}, 483 (1996). 

\bibitem{valla00} T.~Valla, A.V.~Fedorov, P.D.~Johnson, J.~Xue, K.E.~Smith
and F.J.~DiSalvo, Phys.Rev.Lett. {\bf 85}, 4759 (2000);
V.~Vescoli, L.~Degiorgi, H.~Berger and L.~Forro, Phys.Rev.Lett.
{\bf 81}, 453 (1998).  

\bibitem{singley} E.J.Singley {\it et al.} unpublished. 

\bibitem{garoche76} P.~Garoche, J.J.~Veyssie, P.~Manuel and P.~Molinie,
Sol.Stat.Comm. {\bf 19}, 455
(1976). 

\bibitem{finley80} J.J.~Finley and B.S.~Deaver, Sol.Stat.Comm. {\bf 36},
493 (1980). 

\bibitem{thompson72} A.H.~Thompson, F.R.~Gamble and R.F.~Koehler,
Phys.Rev. B {\bf 5}, 2811 (1972). 

\bibitem{su98} X.~Su, F.~Zuo, J.A.~Schlueter, A.M.~Kini
and Jack M. Williams, Phys.Rev. B {\bf 58}, 2944 (1998). 
(resistivity)

\bibitem{kajita87} K.~Kajita, Y.~Nishio, S.~Moriyama, W.~Sasaki,
R.~Kato, H.~Kobayashi and A.~Kobayashi, Sol.Stat.Comm.
{\bf 64}, 1279 (1987). 

\bibitem{su99} X.~Su, F.~Zuo, J.A.~Schlueter, J.M.~Williams, P.G.~Nixon,
R.W.~Winter and G.L.~Gard, Phys.Rev.B {\bf 59},
4376 (1999). 

\bibitem{schwenk83} H.~Schwenk, K.~Andres, F.~Wudl and E.~Aharon-Shalom,
Sol.State Commun. {\bf 45},
767 (1983). 

\bibitem{murata81} K.~Murata, H.~Anzai, G.~Saito, K.~Kajimura and
T.~Ishiguro, J.Phys.Soc.Jpn. {\bf 50},
3529 (1981). 

\bibitem{maeno94} Y.~Maeno, H.~Hashimoto, K.~Yoshida, S.~Nishizaki,
T.~Fujita, J.G.~Bednorz and F.~Lichtenberg, Nature {\bf 372},
532 (1994). 

\bibitem{goldobin98} E.~Goldobin, M.Yu.~Kupriyanov, I.P.~Nevirkovets,
A.V.~Ustinov, M.G.~Blamire and J.E.~Evetts, Phys.Rev.B {\bf 58},
15078 (1998). 

\bibitem{lemberger00} S.J.~Turneaure, T.R.~Lemberger and J.M.~Graybeal,
Phys.Rev.Lett. {\bf 84}, 987 (2000). 

\bibitem{homes93} C.C.~Homes, T.~Timusk, R.~Liang, D.A.~Bonn and
W.N.~Hardy, Phys.Rev.Lett {\bf 71}, 1645 (1993). 

\bibitem{norman} J.E.Hirsch, Physica C {\bf 199}, 305 (1992) and
Phys.Rev.B 62, 14498 (2000); M.R.~Norman, M.~Randeria, B.~Janko
and J.C.~Campuzano, Phys.Rev.B {\bf 61}, 14742 (2000); L.B.Ioffe
and A.J.~Millis, Phys.Rev.B {\bf 61}, 9077 (2000); S.~Chakravarty,
Eur.Phys.J.B {\bf 5}, 337 (1998).


\end{references}
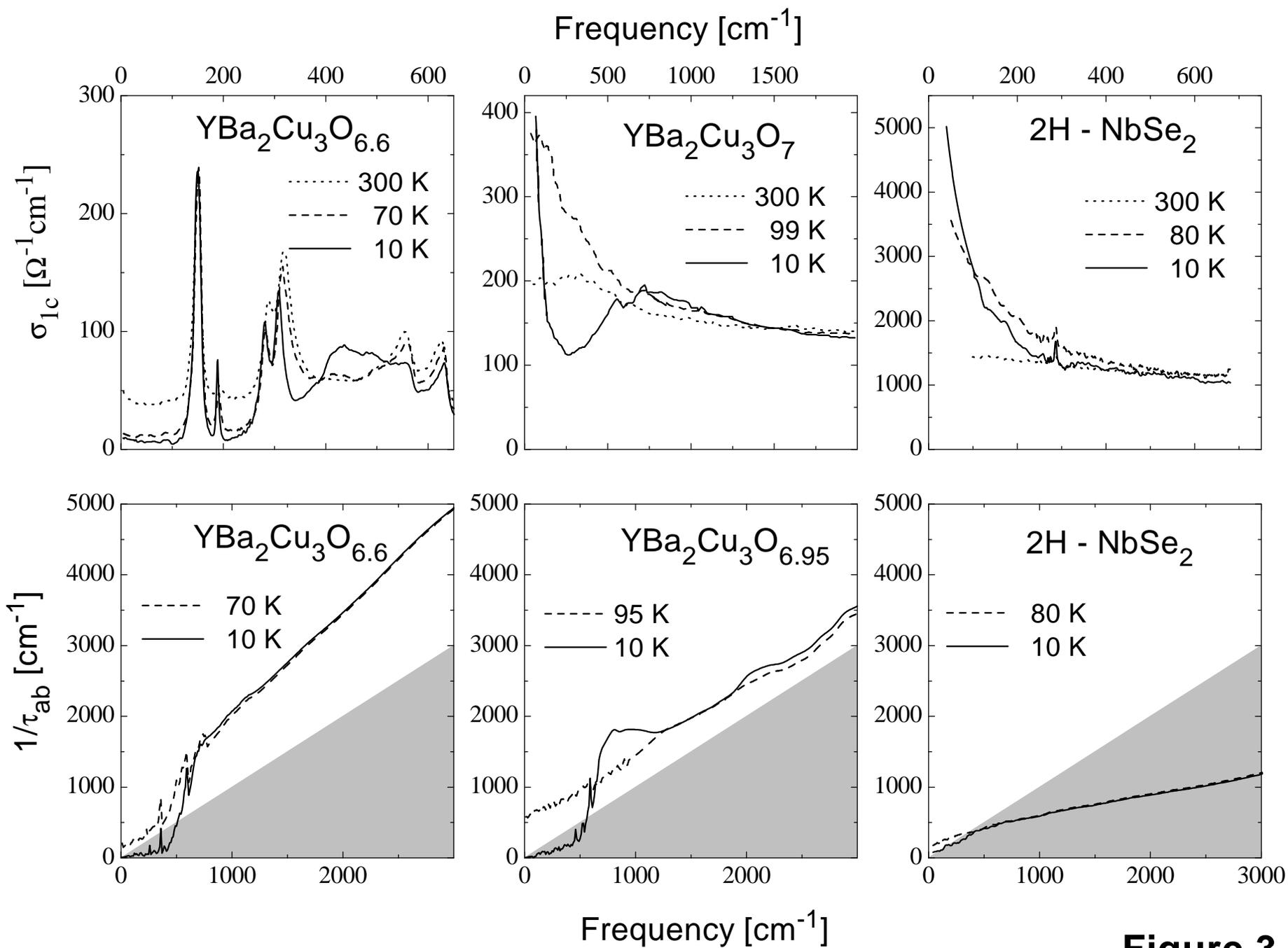
\end{document}